\documentclass[12pt]{article}
\usepackage{amsmath}
\usepackage{amssymb}
\usepackage{psfrag}
\usepackage{epsfig,epsfig,latexsym,amssymb,amsmath}
\usepackage{makeidx}
\usepackage{amsmath}
\usepackage{graphicx}

\usepackage[T1]{fontenc}
\usepackage{lmodern}
\usepackage{listings}

\newcommand{\beq}{\begin{equation}}
\newcommand{\eeq}{\end{equation}}
\newcommand{\beqa}{\begin{eqnarray}}
\newcommand{\eeqa}{\end{eqnarray}}
\newcommand{\ba}{\begin{array}}
\newcommand{\ea}{\end{array}}

\newcommand{\tQ}{\tilde{Q}}

\newcommand{\tG}{\tilde{G}}
\newcommand{\tH}{\tilde{H}}
\newcommand{\tP}{\tilde{P}}

\renewcommand{\beq}{\beqa}
\renewcommand{\eeq}{\eeqa}

\newtheorem{example}{Example}[section]

\begin{document}


\begin{titlepage}

\setcounter{page}{0}

\title{Modified OSD Algorithm with Reduced Gaussian Elimination}

\author{Marc Fossorier, Mahdi Shakiba-Herfeh and Huazi Zhang\\
Huawei Technologies Co. Ltd}

\date{January 2024}

\maketitle
\begin{abstract}

In this paper, the OSD algorithm is modified to perform
a limited GE with $O(N^3 \min\{R, 1-R\}^3)$ complexity for an $(N,K)$ linear block code
of rate $R=K/N$. 

\end{abstract}
\thispagestyle{empty}
\end{titlepage}

\baselineskip 20pt

\section{Introduction}
\label{sec:intro}

Soft decision maximum likelihood decoding (MLD) of linear block codes can be achieved by several
approaches. Some decoders depend on the particular code considered and become code dedicated, while 
others are universal and can be applied to a large class of codes. In this work we are interested 
in this latter class.

Several soft decision decoding approaches are developed in conjunction with the most reliable basis (MRB), which 
is defined as the set of the most reliable independent positions (MRIPs), hence forming an information set. In these 
approaches each considered error pattern is substracted from the hard decison of the MRIPs and the corresponding 
codeword is reconstructed by encoding the corresponding information sequence. The precusor
work of~\cite{dorsch:it74} considers error patterns restricted to the MRB in increasing a-priori likelihood. 
In~\cite{fossorier-lin:it95} the errors patterns are processed in a deterministic order
within families of increasing Hamming weight. This algorithm is referred to as ordered statistic decoding (OSD) 
and the efficiency of these two MRB methods as well as similar ones is compared in~\cite{vf:ieice02}. 

A major drawback of MRB methods is the use of a Gaussian elimination (GE) to put the generator matrix
of the code considered into systematic form with respect to the MRB for encoding the MRIPs. 
For an $(N,K)$ linear block code of rate $R=K/N$, the complexity of GE is $O(N^3 \min\{R,1-R\}^2)$. In this work, we present
an alternate method with a reduced GE in $O(N^3 \min\{R,1-R\}^3)$ complexity.
The main idea is to separate the information positions from the parity positions of the original 
representation of the code in both the MRB and its dual least reliable basis (LRB). As a result the original
information positions also in the MRB do not need to be considered by the GE in G-space decoding. Similarly the
original parity positions also in the LRB do not need to be considered by the GE in H-space decoding. 

\section{Review of OSD}
\label{sec:osd}

The OSD algorithm consists of two main parts~\cite{fossorier-lin:it95}:
(1) construction of the MRB;
and (2) systematic reprocessing of candidate codewords expressed
in the MRB. 

The MRB is constructed as follows:
\begin{itemize}
\item[(a)] 
Order the reliabilities associated with the hard decisions of the received 
values in decreasing order, which defines a first permutation $\lambda_1$. 
\item[(b)]
Apply $\lambda_1$ to the received sequence and to the columns of the generator 
matrix associated with the code considered. Let $\bf{y_1}$ and $G_1$ denote 
the resulting received sequence and generator matrix, respectively, and let
$I_1 = \{0, N-1\}$ be the corresponding indexing. We have
\beqa
|y_{1,i}| \geq |y_{1,j}|
\label{eq:y_ord}
\eeqa
for $i \in I_1, j \in I_1$ and $i < j$. 
\item[(c)]
Perform a GE on $G_1$ from left to right.
The dependent columns found during the GE 
are then permuted after the last independent column found to
obtain a matrix $G_2$ in systematic form, which defines a second permutation
$\lambda_2$. 
\item[(d)]
Apply $\lambda_2$ to the permuted received sequence $\bf{y_1}$, which defines 
a second permuted received sequence $\bf{y_2}$.
\end{itemize} 
For an $(N,K)$ linear block code, the MRB consists of the $K$ positions corresponding 
to the (leftmost) set of independent columns found in step (c).
Note that if $\bf{c_2}$ represents a codeword in the code defined by $G_2$,
then $\lambda_1^{-1}(\lambda_2^{-1}(\bf{c_2}))$ represents the corresponding
codeword in the original code, where $\lambda_j^{-1}$ represents the inverse
permutation to $\lambda_j$.

Once the MRB has been identified, the order-$i$ OSD algorithm
is conducted as follows:
\begin{itemize}
\item[(a)] 
For $0 \leq l \leq i $,
make all possible changes of $l$ of the $K$ most reliable bits in the
hard decision decoding of the permuted noisy received sequence $\bf{y_2}$;
\item[(b)] 
For each change, encode these $K$ bits 
based on $G_2$ (in systematic form), and compute the decoding metric
associated with each constructed codeword; 
\item[(c)] Select the most likely
codeword among the $L(i) = \sum_{l=0}^{i} {{K}\choose{l}}$ constructed
candidate codewords. \end{itemize}
Each value of $l$ determines the reprocessing phase-$l$ of order-$i$
reprocessing. Note finally that an equivalent dual implementation of OSD
based on the processing of the parity check matrix of the code considered
rather than its generator matrix is also 
possible~\cite{fossorier-snyders:it98}.\\

\section{OSD with Reduced GE}
\label{sec:reduced_ge}

\subsection{Generator Matrix}
\label{sec:G_reduced_ge}

Let $G_{REF}$ be a reduced echelon form of the code considered, so that the $K$
columns of identity belong to $G_{REF}$. Let $B_K$ be the location set of the $K$
identity columns of $G_{REF}$ and let $P_{N-K}$ be the location set of the $N-K$ 
parity positions of $G_{REF}$. Note that $G_{REF}$ does not have to be the generator matrix $G$ 
used for encoding but both $G$ and $G_{REF}$ define the same codebook.

For $0 \leq a < b < N$, define $I_1([a,b])$ as the set of consecutives
indeces  $\{a, a+1, \cdots , b\} \subseteq I_1$, $I_1$ being the set of indeces after ordering
defined in Step-(b) of the construction of the MRB in Section~\ref{sec:osd}.

Define:
\beqa
B_{K,MR} & = & B_K \cap I_1([0,K-1]), \nonumber\\
B_{K,LR} & = & B_K \cap I_1([K,N-1]), \nonumber\\
P_{N-K,MR} & = & P_{N-K} \cap I_1([0,K-1]), \nonumber\\
P_{N-K,LR} & = & P_{N-K} \cap I_1([K,N-1]).\nonumber
\label{eq:MRBinter}
\eeqa
We observe
\beqa
|B_{K,LR}| = |P_{N-K,MR}| \leq \min\{K,N-K\},
\label{eq:BKup}
\eeqa

If $\lambda_2$ is the identity function, then $I_1([0,K-1])$ and $I_1([K,N-1])$ correspond to the 
MRB and LRB, respectively.

This separation suggests that in $\lambda_1(G_{REF})$, the GE does not have to be applied to 
$B_{K,MR}$. As a result, the GE is applied to a $K \times (|B_{K,LR}|+N-K)$ submatrix.
The identity part of the obtained systematic form may not correspond to the MRB due
to the dependency occurrences, but the discrepency between the two bases should remain
small. Then the decoding presented in Section~\ref{sec:osd} can be applied.

Alternatively, GE can be applied to only the $|B_{K,LR}|$ rows with all-0 vector
in the MRPs of $B_{K,MR}$. In that case we obtain
\beqa
G^{'}_2 = \left[ \begin{matrix}I_{|B_{K,MR}|}&P_{11}&0&P_{12}\cr
                   0&I_{|B_{K,LR}|} & P_{21}&P_{22}\cr \end{matrix} \right ].
\label{eq:G22}
\eeqa
Note that due the occurrence of dependent columns during the construction of $I_{|B_{K,LR}|}$,
the dimensions of $P_{12}$, $P_{21}$ and $P_{22}$ in~(\ref{eq:G22}) depend on the
number of dependency occurrences. 

For $K < N-K$, based on~\cite{fossorier-snyders:it98} we obtain the parity check matrix
of the code defined by $G^{'}_2$ as
\beqa
H^{'}_2 = \left[ \begin{matrix}Q_{11} & Q_{12} & I_{|P_{N-K,MR}|}& 0 \cr
                               Q_{21} & 0 & Q_{22}& I_{|P_{N-K,LR}|}\cr \end{matrix} \right ].
\label{eq:H22}
\eeqa

\subsection{Decoding Algorithm}
\label{sec:dec_reduced_ge}

Let ${\bf{y^{'}}} = (y^{'}_0, y^{'}_1, \cdots, y^{'}_{N-1})$ be the ordered received sequence
corresponding to $G^{'}_2$ in~(\ref{eq:G22}) and 
${\bf{z^{'}}} = (z^{'}_0, z^{'}_1, \cdots, z^{'}_{N-1})$ the corresponding hard decison
decoding. We decompose
\beqa
{\bf{z^{'}}} & = & ({\bf{z^{'}_{K,MR}}}, {\bf{z^{'}_{K,LR}}})
\label{eq:z'}
\eeqa
with
\beqa
{\bf{z^{'}_{K,MR}}} & = & (z^{'}_0, z^{'}_1, \cdots, z^{'}_{|B_{K,MR}|-1}), \nonumber \\
{\bf{z^{'}_{K,LR}}} & = & (z^{'}_{|B_{K,MR}|}, z^{'}_{|B_{K,MR}|+1}, \cdots, z^{'}_{K-1}). \nonumber
\label{eq:decomp_z'}
\eeqa

Order-0 decoding is achieved as follows:
\begin{itemize}
\item Compute
\beqa
{\bf{c^{'}_{MR}}} & = & {\bf{z^{'}_{K,MR}}} \; \; \left[ I_{|B_{K,MR}|} \;\;\; P_{11} \;\;\; 0 \;\;\; P_{12} \right] \nonumber \\
& = & \left( {\bf{z^{'}_{K,MR}}} \;\;\; {\bf{c^{'}_{11}}} \;\;\; {\bf{0}} \;\;\; {\bf{c^{'}_{12}}} \right).
\label{eq:c01}
\eeqa
\item and
\beqa
{\bf{c^{'}_{LR}}} & = & ({\bf{z^{'}_{K,LR}}} + {\bf{c^{'}_{11}}})  \;\; \left[ 0 \;\;\; I_{|B_{K,LR}|}  \;\;\;  P_{21} \;\;\; P_{22} \right] \nonumber \\
& = & \left( {\bf{0}} \;\;\; {\bf{z^{'}_{K,LR}}} + {\bf{c^{'}_{11}}} \;\;\; {\bf{c^{'}_{21}}} \;\;\; {\bf{c^{'}_{22}}} \right).
\label{eq:c02}
\eeqa
\item Order-0 decoding outputs the codeword
\beqa
{\bf{c^{'}_{0}}} & = & {\bf{c^{'}_{MR}}} + {\bf{c^{'}_{LR}}} \nonumber \\
 & = & \left( {\bf{z^{'}_{K,MR}}} \;\;\; {\bf{z^{'}_{K,LR}}} \;\;\; {\bf{c^{'}_{21}}} \;\;\; {\bf{c^{'}_{12}}}+{\bf{c^{'}_{22}}}\right).
\label{eq:c0}
\eeqa
\end{itemize}

Phase-$j$ reprocessing is achieved based on the $K$-tuples 
\beqa
{\bf{e^{'}}} & = & ({\bf{e^{'}_{K,MR}}}, {\bf{e^{'}_{K,LR}}})
\label{eq:e'}
\eeqa
as error patterns, with $w_H({\bf{e^{'}}}) = j$
and
\beqa
{\bf{e^{'}_{K,MR}}} & = & (e^{'}_0, e^{'}_1, \cdots, e^{'}_{|B_{K,MR}|-1}), \nonumber \\
{\bf{e^{'}_{K,LR}}} & = & (e^{'}_{|B_{K,MR}|}, e^{'}_{|B_{K,MR}|+1}, \cdots, e^{'}_{K-1}). \nonumber
\label{eq:decomp_e'}
\eeqa
\begin{itemize}
\item Compute
\beqa
{\bf{e^{'}_{MR}}} & = & {\bf{e^{'}_{K,MR}}} \; \; \left[ I_{|B_{K,MR}|} \;\;\; P_{11} \;\;\; 0 \;\;\; P_{12} \right] \nonumber \\
& = & \left( {\bf{e^{'}_{K,MR}}} \;\;\; {\bf{e^{'}_{11}}} \;\;\; {\bf{0}} \;\;\; {\bf{e^{'}_{12}}} \right).
\label{eq:ej1}
\eeqa
\item and
\beqa
{\bf{e^{'}_{LR}}} & = & ({\bf{e^{'}_{K,LR}}} + {\bf{e^{'}_{11}}})  \;\; \left[ 0 \;\;\; I_{|B_{K,LR}|}  \;\;\;  P_{21} \;\;\; P_{22} \right] \nonumber \\
& = & \left( {\bf{0}} \;\;\; {\bf{e^{'}_{K,LR}}} + {\bf{e^{'}_{11}}} \;\;\; {\bf{e^{'}_{21}}} \;\;\; {\bf{e^{'}_{22}}} \right).
\label{eq:ej2}
\eeqa
\item Phase-$j$ decoding outputs the codeword
\beqa
{\bf{c^{'}_{j}}} & = & {\bf{c^{'}_{0}}} + {\bf{e^{'}_{MR}}} + {\bf{e^{'}_{LR}}}.
\label{eq:cj}
\eeqa
\end{itemize}
Consequently OSD based on $G^{'}_2$ in~(\ref{eq:G22}) is achieved with a two stage decoding.

OSD based on $H^{'}_2$ in~(\ref{eq:H22}) is also achieved with a two stage decoding based
on the approach described in~\cite{fossorier-snyders:it98}.

\subsection{Complexity}
\label{sec:complexity_red_ge}

To obtain $G^{'}_2$ in~(\ref{eq:G22}),
GE is applied to a $(|B_{K,LR}|) \times (|B_{K,LR}|+N-K)$ matrix, with $|B_{K,LR}|\leq N-K$
from~(\ref{eq:BKup}). Hence its complexity becomes $O((N-K)^3)$.
Similarly $H^{'}_2$ in~(\ref{eq:H22}) is obtained 
with complexity $O(K^3)$. Therefore the complexity of GE in this two stage approach becomes $O(N^3 \min\{R, 1-R\}^3)$.

Based on Section~\ref{sec:dec_reduced_ge}, the complexity to determine 
each codeword is roughly doubled compared to that of the original OSD of~\cite{fossorier-lin:it95}.

\subsection{Generalization}
\label{sec:gen_reduced_ge}

The two stage decoding based on $G^{'}_2$ or $H^{'}_2$ can be generalized into more stages 
to further reduce the complexity of GE. For example the generator matrix corresponding to three stages is
\beqa
G^{'}_3 = \left[ \begin{matrix}I_{|B_{K,MR}|}&P_{11}&P_{12}&0&P_{13}\cr
                   0&I_{|B_{K,LR}|-\alpha} & P_{21}&P_{22}&P_{23}\cr 
                   0&0&I_{\alpha} & P_{31}&P_{32}\cr \end{matrix} \right ],
\label{eq:G23}
\eeqa
with a GE of complexity
\beqa
(N - |B_{K,MR}|) \cdot (|B_{K,LR}|-\alpha) \cdot |B_{K,LR}| + (N - K + \alpha) \cdot \alpha^2.
\label{eq:GE3}
\eeqa

\begin{example}
Consider a (256,128) binary linear code and assume $|B_{K,LR}| = |P_{N-K,MR}| = 64$. 
\begin{itemize}
\item The complexity of full GE is: $256 \times 128 \times 128 = 4,194,304$.
\item The complexity of two stage reduced GE is: $192 \times 64 \times 64 = 786,432$.
\item The complexity of three stage reduced GE with $\alpha=34$ is\footnote{$\alpha=34$ is obtained by a simple optimization of~(\ref{eq:GE3}).}: $192 \times 30 \times 64 + 162 \times 34 \times 34 = 555,912$.
\item The complexity of three stage reduced GE with $\alpha=|B_{K,LR}|/2=32$ is: $192 \times 32 \times 64 + 160 \times 32 \times 32 = 559,992$.
\end{itemize}
\end{example}

We note that this multi stage approach in conjuction with the decoding algorithm of Section~\ref{sec:dec_reduced_ge}
can be applied to the MRB construction as well. In that case the $O(N^3 \min\{R,1-R\}^2)$ complexity of GE can be reduced by a constant.
In~\cite{ma:cl23} a similar structure that also reduces the complexity of GE by a constant is used in conjunction with LC-OSD.

\subsection{Restricted Complexity Decoding Algorithm}
\label{sec:dec_restricted_ge}

Without loss of generality, the matrix $\lambda_1(G_{REF})$ can be rewritten as
\beqa
\tG_2 = \left[ \begin{matrix}I_{|B_{K,MR}|}&\tP_{11}&0&\tP_{12}\cr
                   0& \tP_{21} & I_{|B_{K,LR}|} &\tP_{22}\cr \end{matrix} \right ].
\label{eq:tG22}
\eeqa

For large $N-K$, (\ref{eq:BKup}) implies that the reduced GE may still be quite complex.
To further control the size of the reduced GE, we propose to bound the number of reprocessed rows
to $B_{max}$. Therefore for $|B_{K,LR}| > B_{max}$, $\tG_2$ in~(\ref{eq:tG22}) can be put in the 
following form with column permutations:
\beqa
\tG^{B}_2 = \left[ \begin{matrix}I_{K-B_{max}}&\tP^{B}_{11}&0&\tP^{B}_{12}\cr
                   0& \tP^{B}_{21} & I_{B_{max}} &\tP^{B}_{22}\cr \end{matrix} \right ],
\label{eq:tG22B}
\eeqa
with
\begin{itemize}
\item $\tP^{B}_{11}$: $(K-B_{max}) \times B_{max}$,
\item $\tP^{B}_{12}$: $(K-B_{max}) \times (N-K-B_{max})$,
\item $\tP^{B}_{21}$: $B_{max} \times B_{max}$,
\item $\tP^{B}_{22}$: $B_{max} \times (N-K-B_{max})$.
\end{itemize}

The reduced GE of Section~\ref{sec:G_reduced_ge} can be applied to the last $|B_{max}|$ rows of $\tG^{B}_2$ 
in~(\ref{eq:tG22B}). 
Then the two stage decoding of Section~\ref{sec:dec_reduced_ge} can be performed based on the obtained matrix.

The dual approches for low rate codes can be developed in a straightforward way based on reprocessing in 
H-space~\cite{fossorier-snyders:it98}. The parity check matrix corresponding to~(\ref{eq:tG22B}) is defined by
\beqa
\tH^{B}_2 = \left[ \begin{matrix}\tQ_{11} & I_{|P_{N-K,MR}|} & \tQ_{12} & 0 \cr
                           \tQ_{21} & 0 & \tQ_{22}& I_{|P_{N-K,LR}|}\cr \end{matrix} \right ].
\label{eq:tH22B}
\eeqa

\section{Simulation Results}
\label{sec:sim}

\subsection{BCH(127,113)}
\label{sec:bch127_113}

\begin{figure}[htb]
\begin{center}
\epsfxsize=150mm
\centerline{\epsfbox{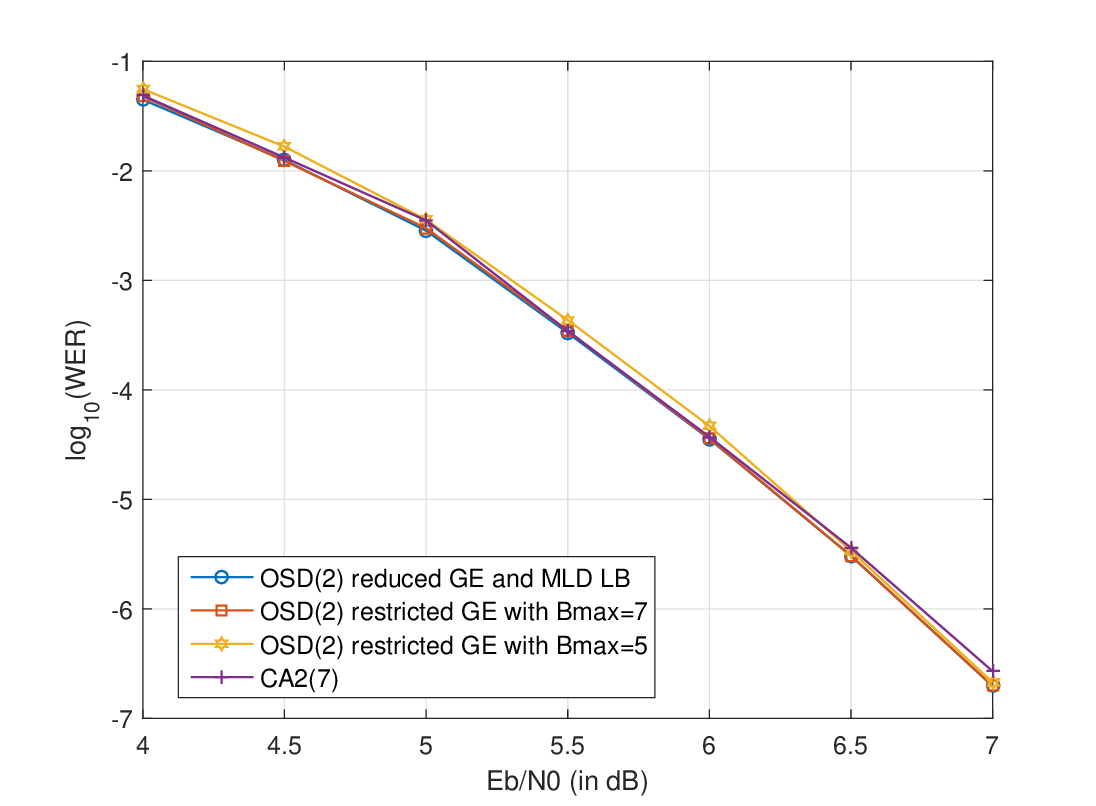}}
\caption{Simulation results for the BCH(127,113) code.}
\label{fig:sim127_113}
\end{center}
\end{figure}

Figure~\ref{fig:sim127_113} compares the word error rate (WER) of the order-2 decoding algorithm with reduced GE
of Section~\ref{sec:dec_reduced_ge}
($N-K = 14$), its restricted complexity versions of Section~\ref{sec:dec_restricted_ge} with $B_{max} = 7$ and 
$B_{max} = 5$ and the Chase algorithm-2 with $p=7$ (referred to as CA2(7)); this algorithm considers 
all $2^7$ test error patterns with support in the $p=7$ LRPs~\cite{chase:it72}. We found 
that all recorded errors for the reduced GE decoding algorithm of Section~\ref{sec:dec_reduced_ge}
are MLD errors, so that these
simulations provide also a lower bound on MLD. At low SNR, all algorithms but that with $B_{max} = 5$ have
their simulated WERs on top of the MLD lower bound; a loss of 0.05~dB is observed for $B_{max} = 5$.
At high SNR, all algorithms but that with $B_{max} = 5$ and CA2(7) have their simulations on top of the MLD 
lower bound; a loss 0.015~dB and 0.055~dB is observed for $B_{max} = 5$ and  CA2(7), respectively.

\subsection{BCH(511,493)}
\label{sec:bch511_493}

\begin{figure}[htb]
\begin{center}
\epsfxsize=150mm
\centerline{\epsfbox{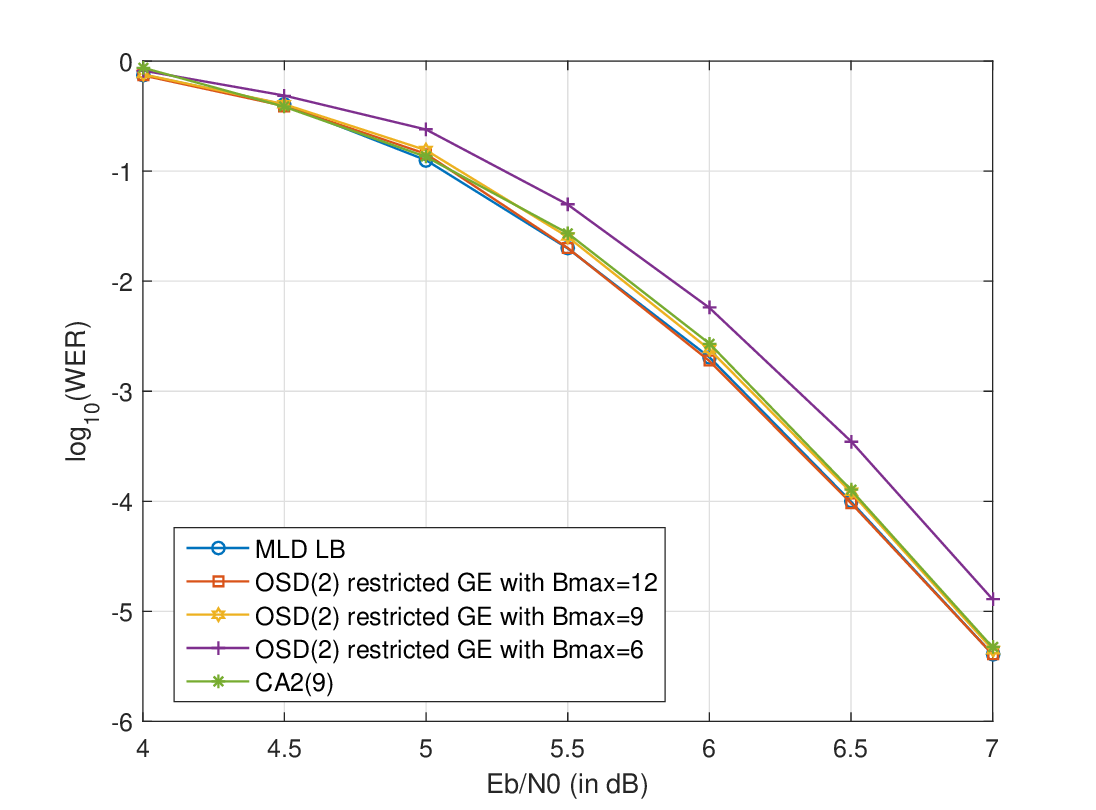}}
\caption{Simulation results for the BCH(511,493) code.}
\label{fig:sim511_493}
\end{center}
\end{figure}

Figure~\ref{fig:sim511_493} compares the WER of the order-2 decoding algorithm 
with restricted complexity of Section~\ref{sec:dec_restricted_ge} for $B_{max} = 6$, $B_{max} = 9$ and
$B_{max} = 12$, the algorithm CA2(9) and a tight lower bound on MLD.
The curve of the algorithm with $B_{max} = 12$ is nearly on top of that of the lower bound on MLD. 
We observe a loss of about 0.05~dB for the algorithm with $B_{max} = 9$ and CA2(9). The loss for the algorithm 
with $B_{max} = 6$ is about 0.25~dB.

\section{Conclusion}
\label{sec:conclusion}

Efficient near MLD decoders for short linear codes have been considered in some ultra reliable low latency communications
(URLLC) scenarios. OSD seems a good candidate due to its good performance and efficient software implementation.
However its hardware implementation is not friendly, mostly due to the GE part. The reasons are:
\begin{itemize}
\item The number of steps/cycles is not deterministic and subjet to the dependency occurrences when constructing the MRB.
\item Sorting a large number of values is not area efficient.
\item GE involves cubic complexity and quadratic memory.
\end{itemize}
Consequently these results achieve savings for both software and hardware implementations with good performance trade-offs.

\end{document}